Resource-Frugal Classification and Analysis of Pathology Slides Using Image Entropy

Steven J. Frank

Abstract—Pathology slides of lung malignancies are classified using resource-frugal convolution neural networks (CNNs) that may be deployed on mobile devices. In particular, the challenging task of distinguishing adenocarcinoma (LUAD) and squamous-cell carcinoma (LUSC) lung cancer subtypes is approached in two stages. First, whole-slide histopathology images are downsampled to a size too large for CNN analysis but large enough to retain key anatomic detail. The downsampled images are decomposed into smaller square tiles, which are sifted based on their image entropies. A lightweight CNN produces tile-level classifications that are aggregated to classify the slide. The resulting accuracies are comparable to those obtained with much more complex CNNs and larger training sets. To allow clinicians to visually assess the basis for the classification — that is, to see the image regions that underlie it — color-coded probability maps are created by overlapping tiles and averaging the tile-level probabilities at a pixel level.

1.      Introduction

"Deep learning" approaches have been applied to a wide range of medical images with the objective of improving diagnostic accuracy and clinical practice. Many efforts have focused on images that are inherently small enough to be processed by convolutional neural networks (CNNs), or which can be downsampled without loss of fine features necessary to the classification task [1–6]. In general, CNNs perform best at image sizes below 600 × 600 pixels; larger images entail complex architectures that are difficult to train, execute slowly, and require significant memory resources. Among the most challenging medical images to analyze computationally are digital whole-slide histopathology images, which are often quite large — 10,000 to more than 100,000 pixels in each dimension. Their large size means that even traditional visual inspection by trained pathologists is difficult.

To make such images amenable to CNN analysis, researchers have decomposed them into much smaller tiles that are processed individually. A probability framework is applied to the tile-level classifications to classify the slide. Khosravi et al. [6], for example, utilize tiles of 224 × 224 and 229 × 229 pixels, which is typical for the complex CNN architectures that have been employed to date. Coudray et al. [3] use 512 × 512 tiles. Such tiles cover an extremely small region of the original slide. For a typical 1 Gigabyte (GB) slide image, even the larger tiles used by Coudray et al. [3] each cover only 0.026% of the slide area. Not all tiles are equally meaningful for classification. Coudray et al. [3] exclude tiles having backgrounds exceeding 50% of the image, while Yu et al. [1] take a staged approach. First they decompose a whole-slide image into non-overlapping 1000 × 1000 pixel squares and select the 200 densest squares, then further subdivide the selected squares into tiles of unspecified sizes (but presumably comparable to those of Khosravi et al. [6] since similar CNN architectures are used).

The most successful recent studies have achieved performance comparable to that of experienced pathologists. Among the most challenging tasks is distinguishing LUAD from LUSC tissue. Coudray et al. [3] obtained an AUC value of 0.950 for LUAD/LUSC classification when probability aggregation is based on averaging of classification probabilities. Yu et al. [1] report AUC values of 0.883-0.932. These studies have involved large datasets and complex CNN architectures. Coudray et al. [3] utilize the Inception v3 architecture, which has 48 convolutional layers; at their tile size of 512 × 512 pixels, an Inception v3 CNN is about 1.8 GB in size. Yu et al. [1] utilize several architectures: VGGNet-16, Residual Network-50 (ResNet), GoogLeNet, and AlexNet. The smallest of these, AlexNet, has only eight functional layers but nonetheless imposes a high bandwidth burden: its design involves 60 million parameters and requires over 729 million floating-point operations (FLOPs) to classify a single image [7]. Coudray's [3] dataset contained 567 LUAD slides and 608 LUSC slides, while Yu et al. [1] employed 427 LUAD slides and 457 LUSC slides. Another study [8] of lung adenocarcinoma slides used 422 whole slide LUAD images obtained from a single source.

A longstanding impediment to clinical adoption of machine-learning techniques is the inability of many such techniques to convey the rationale behind a classification, diagnosis or other output [9]. Black-box models whose reasoning is opaque or impervious to retrospective analysis may pose clinical dangers that outweigh the benefits of a computational approach [10]. Until recently, CNNs have fallen squarely within the black-box category, but techniques such as gradient-weighted class activation maps ("Grad-CAM") [11] have pried the box open, highlighting the image regions important to a CNN classification. Yu et al. [1], for example, show Grad-CAM images of individual tiles processed by their system.

While the ability to visualize regions of an image important to classification is useful, it does not necessarily address clinical acceptance. Yu's [1] Grad-CAM images each represent perhaps 0.0045% of the total slide area. Painstaking analysis of many such images could help validate the proposition that the CNN is "looking" where it should. But for any given slide classification, Grad-CAM cannot realistically illuminate its underlying basis; the Grad-CAM images are too small and a readable map of them superimposed on the slide would be impossibly large. Moreover, identifying which image regions attract the attention of a CNN does not reveal the underlying rationale for a classification — only the pixels on which the classification, whatever its basis, depended most strongly.

The computational demands of CNNs pose an additional challenge to their widespread deployment, particularly given the trend toward telemedicine and increasing clinical use of mobile devices [12]. Although the computational capacity of tablets and phones continues to grow — today's devices typically include graphics processing units and high-end devices may feature dedicated neural processing units — the processor typically runs many background and foreground tasks that compete for cycles and battery life. Indeed, compute-intensive applications can increase a device's thermal profile and trigger throttling, which slows computation altogether. In parallel with efforts to augment hardware capacity, researchers have attempted to reduce the memory and energy burden imposed by CNNs. One promising approach uses quantized representations of internal CNN values to make them more compact. Quantization may be employed to make CNN training tractable in constrained execution environments [13] or to simplify processing of already-trained models [14,15]. Practical medical deployments will involve trained CNNs, quantization of which has been shown to accelerate inference and reduce memory footprint in some applications without adversely affecting classification performance [16].

Here, rather than simplifying or accelerating execution of a complex model, our strategy is to adapt simple CNN architectures to difficult classification tasks using staged image resizing and visually based data sifting. Like Yu et al. [1], we first downsample whole-slide images but to a much larger intermediate size that preserves visible anatomic characteristics. These intermediate images are decomposed into smaller square tiles of varying sizes that can be evaluated individually. The tiles are sifted using image entropy as a visual criterion. These steps — initial image rescaling followed by trials at multiple tile sizes — provide two "knobs" for optimization. The initial rescaling must preserve sufficient anatomic detail to support accurate classification. Previous work [17] suggests that one or two tile sizes will emerge as optimal for a given classification task, a result observed here as well. Clear winners emerged when the best CNN models were tested with out-of-sample tiles. A model's performance on the out-of-sample set reflects its generalizability.

CNNs for even the largest tile sizes tested are lightweight enough to be deployable on mobile devices such as phones and tablets. To permit visualization on such devices of the basis for classifying an input image, color-coded probability maps are created; in particular, tiles that survived sifting are overlapped and their tile-level probabilities averaged at a pixel level. Unlike Grad-CAM images, these maps show the distribution and intensity of the actual predictions across the image.

2. Materials and Methods

*2.1 Whole slides*

Forty-two LUAD slides and 42 LUSC cell carcinoma slides were downloaded from the GDC portal of the National Cancer Institute. These averaged about 1 GB in size and contained varying amounts of empty, non-image area. The pulmonary pathologists' evaluations from the TCGA study were used as the ground truth for classification.

*2.2 Tile generation*

The whole-slide images were first rescaled such that the longer dimension of the rescaled image did not exceed 6000 pixels. This size seemed appropriate the diagnostically significant anatomy, a result confirmed by subsequent testing with images rescaled to maximum dimensions of 4500 and 8000. The result was an average image size of 8.5 Megapixels (MP) and a maximum size of 13.5 MP. (The maximum possible image size would be 36 MP, or 6000 × 6000 pixels.) Tiles were prepared from these rescaled images.

For comparative purposes, an effort was made to adhere as much as practicable to the data-preparation techniques used by Coudray et al. [3]. This resulted in two tile sets generated using different methodologies: one set was obtained using image entropy as a sifting criterion, and the other excluded majority-background tiles.

The entropy criterion attempts to identify the most visually salient portions of an image by excluding tiles whose individual image entropies fall below the entropy of the entire image. As described in [17], image entropy represents the degree of nonredundant information in an image and is given by:

$$H = -\sum_{k} p_k \log_2(p_k)$$

where $p_k$ is the probability associated with each possible data value $k$. For a two-dimensional eight-bit grayscale image (or one channel of a color image), $k$ spans the 256 possible pixel values [0..255]. Because entropy does not scale with image size, even small images with high visual diversity can have large entropies, and it is meaningful to compare the entropy of a tile with that of the larger image from which it is drawn. Each rescaled image was decomposed into overlapping tiles (i.e., the tiles intercepted overlapping image regions) that collectively, and redundantly, covered the entire image. Tiles with entropies below that of the full image were excluded, resulting in an average retention ratio of 0.4145 — i.e., about 41% of the generated tiles survived sifting. We refer to the retained tiles as "Entropy tiles."

Coudray et al. [3] exclude low-information tiles based on the proportion of background; in particular, they identify and discard tiles for which a majority of pixel values fall below 220 in the RGB color space. However, applying this selection criterion to our dataset resulted in no tile exclusion at all. Presumably this retention ratio of 1.0 resulted from varying degrees of discoloration in the slide backgrounds. To exclude majority-background tiles as proposed by Coudray et al. [3], we created 8-bit grayscale tile counterparts and used these to determine which of the original (RGB) tiles to retain. In particular, we excluded tiles for which the values of a majority of grayscale pixels exceeded 240 (nearly white) or fell below 15 (since a few tiles were completely or mostly black). We refer to these as "Threshold Gray" tiles. The average retention ratio for these tiles was 0.560. The retention ratios for Entropy and Threshold Gray tiles were substantially consistent across tile sizes.

To obtain similar numbers of tiles of each type and produce enough larger tiles for training and testing purposes, we overlapped them to different degrees. Tile populations and overlap amounts are set forth in Supplementary Table 1. Test and training sets were prepared using a 70/30 training/test split, corresponding to 58 training images (29 LUAD, 29 LUSC) and 26 test images (13 LUAD, 13 LUSC). To ensure that the observed performance would not arise due to this training/test set partition, two more partitions were created so that no image appeared in more than one test set. This corresponds roughly to stratified three-fold cross-validation. The three partitions are labeled XValSet1, XValSet2, and XValSet3 in Supplementary Table 1.

*2.3 Deep learning model*

The CNN architecture employed in this study was selected to minimize the number of convolutional layers and consequent trainable parameter count. Three dropout layers mitigate the risk of overfitting to the small dataset employed in this study. Using the architecture shown in Fig. 1, we trained for 35 epochs in each training/test partition using a batch size of 16, a binary cross-entropy loss function, an Adam optimizer, and random horizontal and vertical flip data augmentation. More significant data augmentation resulted from the high degree of tile overlap employed (up to 75% as detailed in Supplementary Table 1).

After each training epoch, the resulting model was saved. With a decision boundary at 0.5, higher prediction probabilities indicated a LUSC classification and lower probabilities corresponded to LUAD. Tile-level prediction probabilities were averaged to produce an image classification. As noted in [17], averaging across tiles renders tile-level validation largely irrelevant to overall classification accuracy, so early stopping is not a useful strategy, nor are validation metrics particularly meaningful. Instead, optimal practice is to train for a sufficient number of epochs to obtain a distinct accuracy peak.

Source code for this model has been posted.[1]

---

[1] https://github.com/stevenjayfrank/A-Eye.

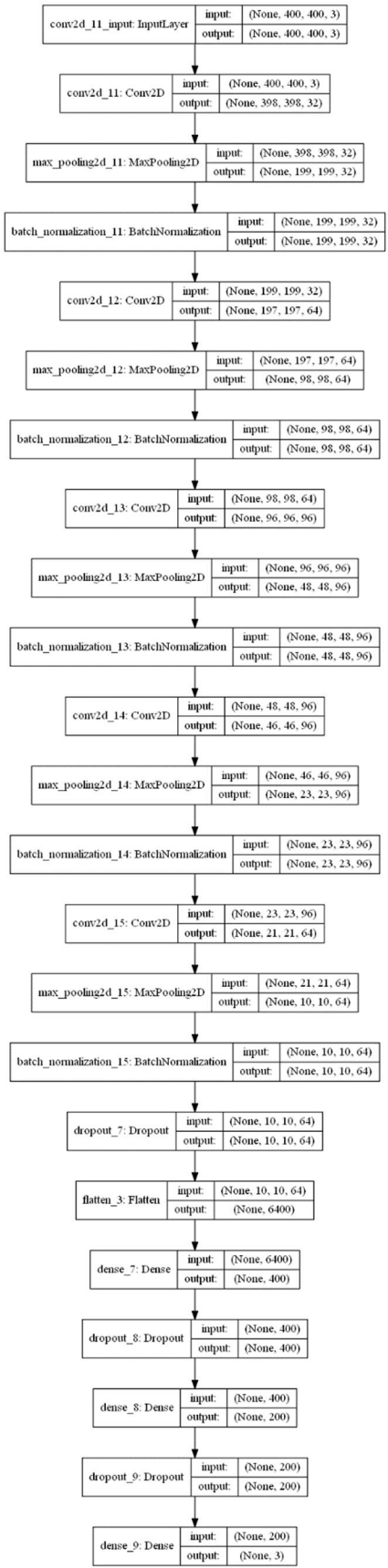

Figure 1 – Convolution neural network architecture employed for this study

## 3. Results

The purpose of this study was to explore whether small CNN architectures could be adapted for difficult classification tasks on very large medical images. A deliberately small dataset was used in order to "stress test" the different tile-sifting criteria and reveal performance differences among them. The small dataset also permits an out-of-sample test to be readily defined, as described below, and helps us assess whether acceptable model performance can be attained even when data is limited.

Each of the three training/test partitions included Entropy tiles derived from 26 whole slides (13 LUAD, 13 LUSC tissue) that were not present in any other test partition. For each partition, tiles derived from the remaining 58 whole slides were used to train CNN models over 35 epochs as described above. This procedure was repeated for each tile size. Within a partition, all 35 resulting models generated for each tile size were tested against the partition test set. As shown in Supplementary Table 1, the high degree of tile overlap permitted training based on at least tens of thousands of tiles, and testing based on at least thousands of tiles.

The results of testing appear in Supplementary Table 2. Accuracies were computed as a straight fraction of correctly classified test images. Average accuracies over all tile sizes were close to 90% across the three partitions and variation among the partitions was limited: the high/low range never exceeded 15%, with an average range of 9.4%. The average prediction probability variance was low (about 6% on average across all partitions and tile sizes) and exhibited very little change among the different tile sizes tested.

The first partition (XValSet1) produced CNN models exhibiting the highest absolute prediction accuracies (96% for three tile sizes). As noted, an image classification reflects the average of the tile-level prediction probabilities for that image. It was observed that most of the average probabilities were much closer to the probability limits of 0 and 1.0 than to the decision boundary of 0.5. Classification by majority vote (i.e., based on agreement of tile-level predictions rather than their probability average) produced identical results.

To assess the benefits of training with Entropy tiles, Threshold Gray tiles derived from images in the first partition were generated and the training and testing procedures repeated. Once again, all 35 models generated during training for each Threshold Gray tile size were tested against the partition test set. Fig. 2 illustrates the comparative accuracies obtained for Entropy and Threshold Gray tiles derived from the first training/test partition. The average prediction probability variance was higher (about 11% on average across tile sizes) than for Entropy tiles and was spread over a fairly wide range (7.6% to 13.9%). This suggests greater prediction stability for Entropy tiles, likely due to a consistently smaller amount of irrelevant visual information that can act as prediction-degrading noise — a possibility consistent with the much lower proportion of tiles retained using the Entropy sifting criterion (41%) as compared with Threshold Gray (56%).

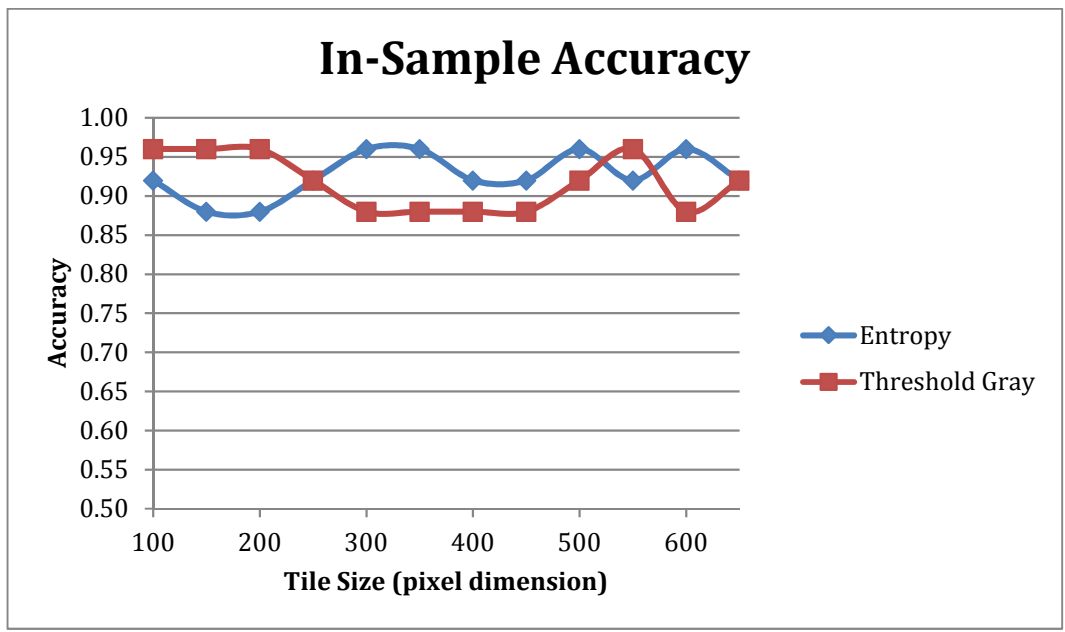

Figure 2 – Comparative classification accuracies for tiles sifted with Entropy and Threshold Gray image criteria. In each case the tiles were derived from the first training/test partition. Tile sizes ranged from $100 \times 100$ to $650 \times 650$ pixels in steps of 50.

The plot shows that models trained using Entropy tiles performed similarly to those trained with Threshold Gray tiles, although the results differed slightly across most tile sizes. In both cases, at least some tile sizes produced models with prediction accuracies of 96%, i.e., 25/26 images correctly classified. The high overall accuracies reflect the amount of classification-supporting image information contained even in the smallest tiles. Even at a maximum rescaled image size of 6000 × 6000 pixels, any tile covers 27.78 times the slide area that it would in the absence of rescaling, assuming a 1 GB slide. As a result, much more of the anatomy appears in a tile and enters each round of CNN processing.

The best-performing models generally appeared after 20 to 30 epochs of training, following which overfitting began to set in. Discriminating among the better models can be elusive. Most simply, the classification margin — i.e., the average distance between an image-level classification probability and the decision boundary — can be readily computed. This metric can be misleading, however, owing to the ensemble effects of averaging across tiles, and does not correlate well with classification accuracy. A better test would utilize out-of-sample slides, ideally with some distinguishing characteristic from the in-sample test slides.

TGCA slides originate from different sources, which are specified by a "TSS" (tissue source site) field in the title code assigned to each slide. Although tissue-preparation techniques are standard, an out-of-sample test set was assembled using 12 slides (six LUAD, six LUSC) originating mostly with sources other than those used for training and testing, and with visual characteristics (such as overall sparseness) that seemed unusual relative to the training and test sets. (All slides in the primary and out-of-sample datasets are identified in Supplementary Table 3.) These slides were rescaled, tiled, and sifted into separate sets using the Entropy and Threshold Gray criteria as described above. As shown in Fig. 3, testing with these tile sets resulted in substantially more variation and lower overall accuracies. It also better revealed performance differences between tile-sifting criteria and suggested optimal tile sizes. Fig. 3 also shows the poor classification accuracies obtained using the entire set of unsifted tiles, demonstrating the distortive effects of excessive background regions within tiles.

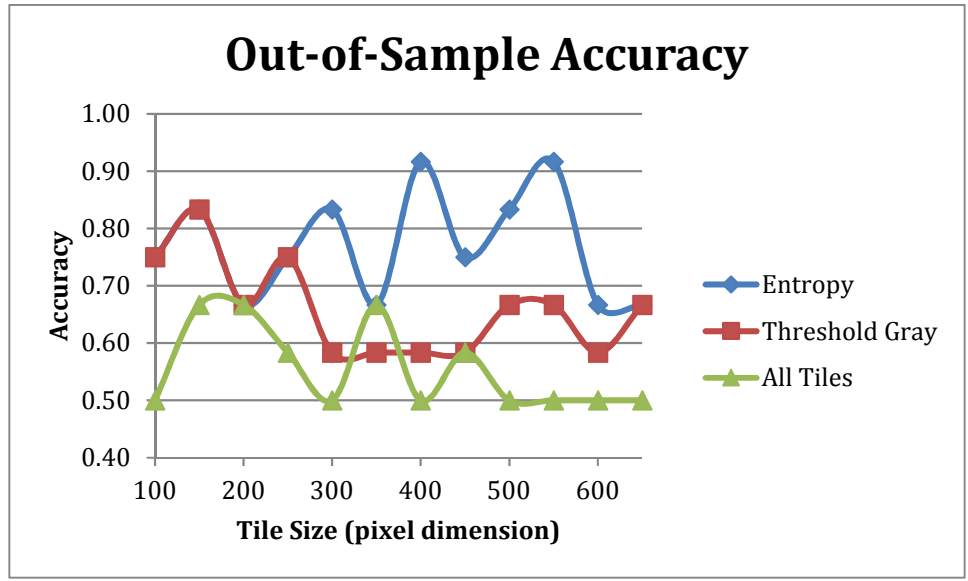

Figure 3 – Out-of-sample testing results using Entropy and Threshold Gray tiles, with comparison to unsifted tiles. Most were drawn from slides originating with tissue sources different from those that supplied the in-sample training and test images.

The out-of-sample results conform more closely to experience with other classification tasks. As noted above, it is typical for one or two tile sizes within a range to exhibit superior performance. At these tile sizes, critical anatomical detail is present at an optimal density and resolution. In addition, the characteristic pattern of disease may involve unaffected tissue (and therefore favor a larger tile size); for example, some features strongly prognostic of breast-cancer survival have been found to lie in the region surrounding tumor cells.[18] The best-performing tile sizes represent a highly disease-specific balance among different (and possibly competing) visual criteria.

The out-of-sample results show a clear performance advantage for Entropy tiles over Threshold Gray tiles at most tile sizes, and higher maximum accuracy: 92% (11/12 slides classified correctly) compared to 83% (10/12 correct classifications). The performance variation among tile sizes accords with the expected pattern. In this case, best results were obtained using $400 \times 400$ and $550 \times 550$ pixel tiles. The much smaller degree of variation among in-sample tile sizes may reflect excessive homogeneity within the primary dataset; this is consistent with the very similar average variance levels among tile sizes noted above.  Overfitting is a less likely cause given the out-of-sample performance variation among the same models that gave rise to the results shown in Fig. 2.

The disadvantage of Entropy tiles is the computational complexity of their production. The time required to sift a set of tiles using the Entropy criterion was found to exceed that needed for Threshold Gray tiles by a factor of 50, and the computing time per tile scales linearly with the pixel count.  That difference may as a practical matter preclude deployment of Entropy-based systems on mobile devices.  However, we found that once the models are trained using Entropy tiles, their classification performance on Threshold Gray candidate tiles more closely tracks the results obtained with Entropy test tiles.  This is illustrated in Fig. 4.

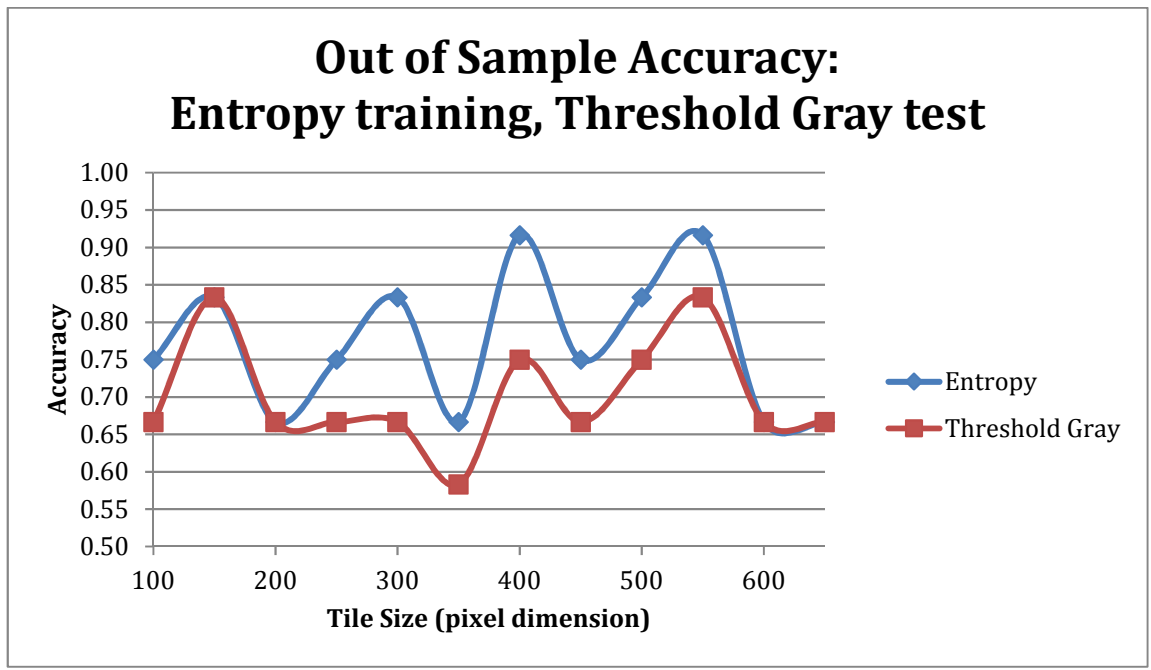

Figure 4 – Out-of-sample results using models trained using Entropy tiles.  Candidate tiles derived from the same slides were sifted using the Entropy and Threshold Gray criteria.

Still, while the performance patterns are similar, the absolute differences in accuracy are not insignificant; Entropy sifting produces a consistent advantage.  But the out-of-sample test was designed to emphasize such differences.  In a real deployment, the training sets would ideally generalize to any source and across morphological variations.  We would expect the performance differences to diminish across tile sizes and likely between Entropy and Threshold Gray tiles as well.  For visualization as described below, smaller tiles produce higher-resolution probability maps, and may be preferred for this reason over larger tiles that produce similar classification accuracies.

4.       Visualization

    Grad-CAM uses the "feature maps" produced by a convolutional layer (typically the final one).  Class activation maps (CAMs) are produced by projecting back the weights of the output layer onto the convolutional feature maps.  This has the effect of ranking the image regions by their importance to the classification.  Grad-CAM generalizes this technique to essentially any CNN architecture.  While the CAM and Grad-CAM techniques have lifted the veil to reveal the basis for CNN classification decisions, Fig. 5 illustrates their shortcomings for medical images.

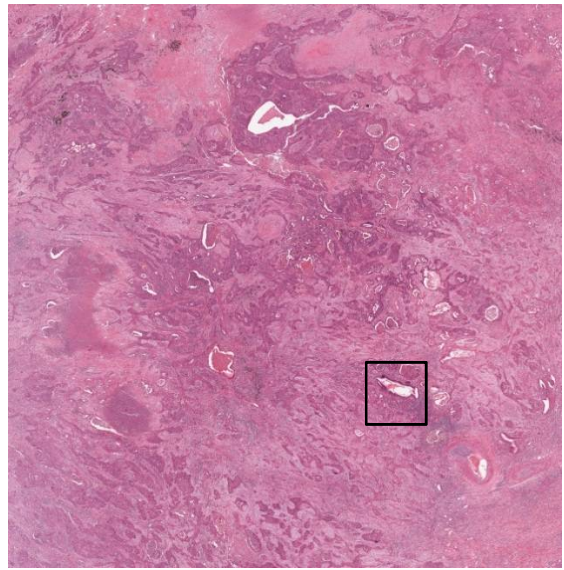

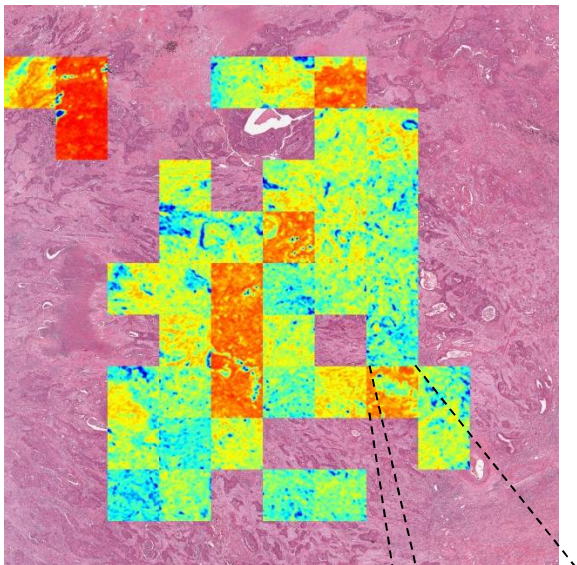 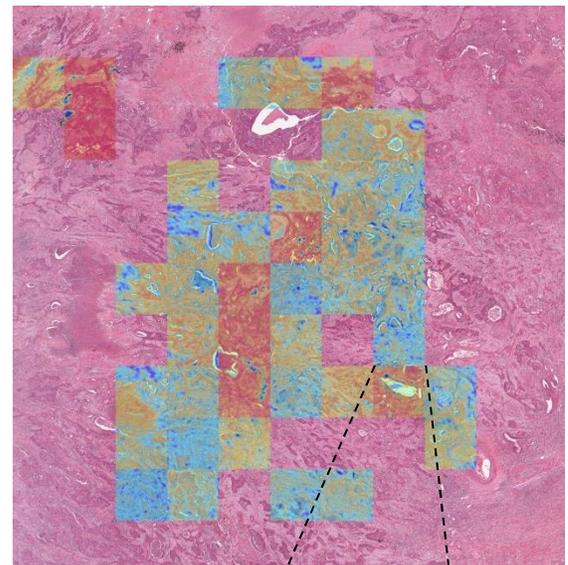

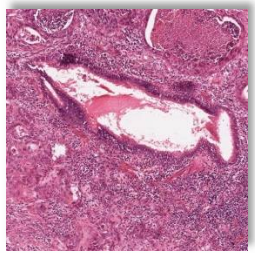 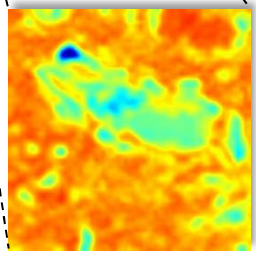 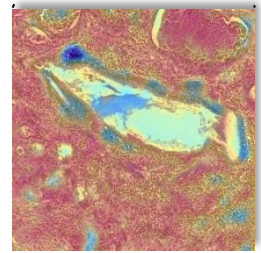

Figure 5: Grad-CAM "heat maps" indicate the the relative importance of different regions within the tile to the classification decision. (a) Resized histopathology image of LUSC lung tissue; (b) Grad-CAM maps generated for non-overlapping $550 \times 550$-pixel Entropy tiles of resized image; (c) same maps superimposed on resized image with partial transparency; (d) single $550 \times 550$ pixel tile from resized image, the origin of which is indicated in (a); (e) its Grad-CAM heat map; and (f) heat map superimposed with partial transparency.

As noted earlier, Grad-CAM heat maps must be produced at the level of a tile covering a very small region of the image. While the tile-level Grad-CAM heat map shown in Fig. 5(f) is illuminating for the particular tile, overall image classification is based on many tiles as illustrated in Fig. 5(c). Examining these individually to assess whether the CNN is focusing on the correct anatomic features would be quite laborious. The Grad-CAM tiles cover only a discontiguous minority of the overall image because only a minority of tiles survive sifting. Much more of the image is covered if tiles are overlapped, but overlapping is not an option for Grad-CAM heat maps since each contains unique information.

As an alternative, we create probability maps for visualizing of classification probabilities at a subimage level (Fig. 6). The probability map color-codes the probabilities assigned to the examined regions of an image. Using a high degree (92%) of overlap results in coverage of large, contiguous regions of the image and fine visual features. Each colored pixel represents the classification probability averaged over all sifted tiles containing that pixel. Thus, the probability map represents the union of all pixels included in any tile.

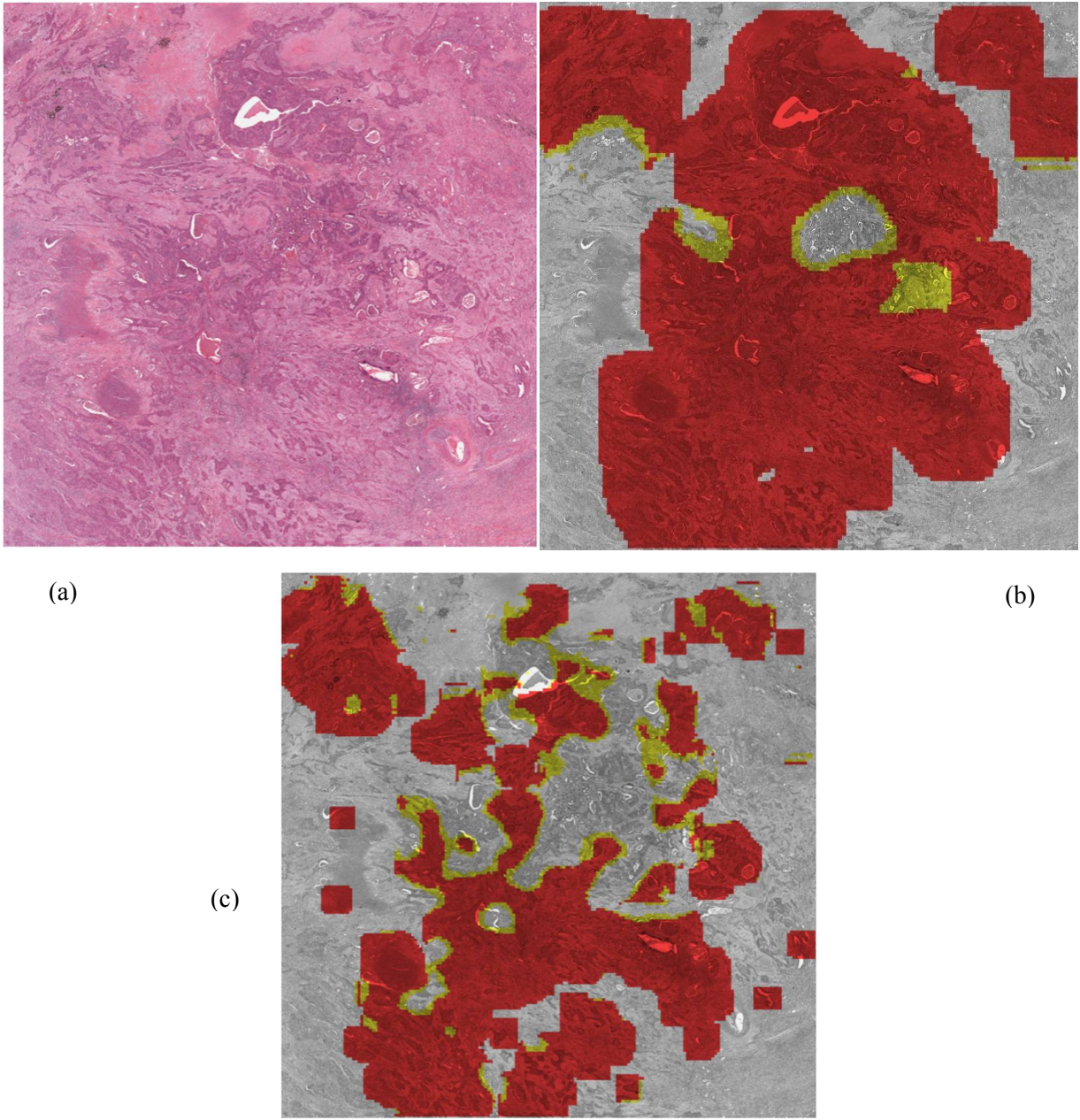

Figure 6: (a) resized histopathology image of LUSC lung tissue; (b) color-coded probability map of the classified regions using 550 × 550 pixel Entropy tiles; and (c) using 250 × 250 pixel Entropy tiles. With a decision boundary at 0.5, red corresponds to high-likelihood ($p \geq 0.65$) classification as squamous cell carcinoma, and gold to moderate-likelihood ($0.5 \leq p < 0.65$) classification as squamous cell carcinoma. Misclassified and unclassified regions are not colored since they do not contribute to the classification.

As demonstrated in Fig. 6, smaller tiles produce finer features. Ultimately, however, the optimal tile size for creating probability maps is the one that delivers the best classification accuracy, not the highest feature resolution. The apparent precision of a fine-featured map generated with a mediocre model may be very misleading. Larger tiles produce clinically "safer" probability maps in the sense that the critical anatomy will likely be colored (at the price of extraneous coloring). Unlike Grad-CAM heat maps, probability maps do not visually rank image regions in terms of classification importance; instead, they highlight all regions relevant to the classification and visually reveal the associated probability levels. This permits clinicians to readily determine whether a classification is based on the proper anatomy. Conversely, it can help focus a clinician's attention on key slide regions to make a non-automated diagnosis. If a training set is large enough to span the full range of morphologies encountered by pathologists, variegated probability maps may also provide a useful guide to the subtleties underlying a classification (highlighting anatomic regions not conventionally associated with a diagnosis, for example) and allow clinicians to identify possible sources of error.

5.    Discussion

The foregoing results suggest that lightweight CNNs can assist in making diagnoses based on very large histopathology images. The strategy of intermediate image scaling followed by training and testing over multiple tile sizes enabled simple CNN models to successfully distinguish difficult-to-classify lung cancer types. Particularly when pre-trained on Entropy tiles, the accuracies were equivalent to those obtained with much larger architectures. Probability maps enable ready visualization of the basis for a classification.

The architectures and the working images employed in this study are small both in size and the computational load they impose. For the best-performing tiles sizes of $400 \times 400$ and $550 \times 550$ pixels, the corresponding five-layer CNN models are only 33.5 MB and 97.2 MB in size, respectively. Until recently, Apple Corp. limited the size of executable applications ("apps") for its mobile devices to 200 Megabytes (MB) in deference to telecommunication carrier concerns over bandwidth usage. Although the 200 MB limit is no longer mandatory, the proliferation of business and personal apps competing for still-finite storage space continues to favor memory-frugal deployments. Since the CNN will occupy most of the required storage space for a diagnostic application as described above, keeping its size below 100 MB should ensure the app will qualify as "lightweight."

The number of parameters associated with a CNN refers to the weights learned during training, i.e., the operative mathematical elements that produce a prediction. The size of this number reflects the complexity of a CNN and its computational burden [19]. FLOPs reflect the number of operations are required to run a single instance of a given model. The commonly used ResNet50 architecture uses 23 million parameters and 4 billion FLOPs to process $224 \times 224$ images; Inception v3, with 48 convolutional layers, uses 24 million parameters and 6 billion FLOPs. By contrast, our five-layer CNN model uses just 597,000 parameters and 934 million FLOPs at the same image size.[2] Even with the much larger $550 \times 550$ images described here, the five-layer model uses only 8 million parameters and 1.8 billion FLOPs.

Both the small models and the intermediate-scale images used in this study can readily be downloaded and run on a mobile device. Images rescaled as described above have a maximum size of only 36 MB; it should be possible to decompose such images into Threshold Gray tiles and process them on a mobile device within an acceptably short time period. It should also be possible to create probability maps on the device.

Various useful clinical scenarios can be envisioned. A pathologist's tablet might, for example, receive a rescaled histology image from a microscopy device or server via Bluetooth, WiFi or other wireless protocol. Importantly, it is not essential for the pathologist to be physically proximate to the image source, and the image is small enough that transmission is essentially instantaneous. The pathologist could select one of several tablet-stored CNN models and launch an on-board application that decomposes the histology images into tiles, presents them to the selected CNN model, and reports the classification. The basis for the classification may be displayed in the form of a probability map. Should the classification appear questionable, pathologist could select (or download from a larger repository) another CNN model; once again, the small model size facilitates fast downloading (as well as updates, which could be pushed out to user devices automatically). Such scenarios give pathologists ready access to a variety of diagnostic tools. Freed from the equipment used to obtain the initial images and the need for powerful workstations to analyze them, they can more easily pursue telemedicine collaborations with distant colleagues.

---

[2] The FLOPs figure is comparable to that of the MobileNet v2 architecture, but even that more compact architecture, designed expressly for mobile devices, uses 4.3 million parameters.

Further work is needed with larger datasets and different diagnostic tasks to establish suitability of our approach for clinical use, as well as to explore extensions. In the case of solid tumor masses and other gross anatomic features, for example, probability maps may provide a rough tissue segmentation partitioning the diseased region. Fusing probability maps generated from multi-modal images may improve segmentation quality [20] or resolution, or highlight different diagnostically significant regions. It should also be stressed that the binary classification task described above performs best on images consisting mostly of malignant tissue, or where normal tissue is (erroneously) classified as one or the other type of malignancy with equal probability. Outside these conditions, the classification task is ternary and should be approached as such, e.g., using a softmax activation function and training slides that have been segmented by experts.

It is hoped that resource-frugal implementations of deep-learning algorithms will help enhance diagnostic accuracy, advance clinical collaborations, and ultimately improve patient outcomes.

References

[1]  K.H. Yu, F. Wang, G.J. Berry, C. Ré, R.B. Altman, M. Snyder, I.S. Kohane, Classifying non-small cell lung cancer types and transcriptomic subtypes using convolutional neural networks, J. Am. Med. Informatics Assoc. (2020). https://doi.org/10.1093/jamia/ocz230.

[2]  B.E. Bejnordi, M. Veta, P.J. Van Diest, B. Van Ginneken, N. Karssemeijer, G. Litjens, J.A.W.M. Van Der Laak, M. Hermsen, Q.F. Manson, M. Balkenhol, O. Geessink, N. Stathonikos, M.C.R.F. Van Dijk, P. Bult, F. Beca, A.H. Beck, D. Wang, A. Khosla, R. Gargeya, H. Irshad, A. Zhong, Q. Dou, Q. Li, H. Chen, H.J. Lin, P.A. Heng, C. Haß, E. Bruni, Q. Wong, U. Halici, M.Ü. Öner, R. Cetin-Atalay, M. Berseth, V. Khvatkov, A. Vylegzhanin, O. Kraus, M. Shaban, N. Rajpoot, R. Awan, K. Sirinukunwattana, T. Qaiser, Y.W. Tsang, D. Tellez, J. Annuscheit, P. Hufnagl, M. Valkonen, K. Kartasalo, L. Latonen, P. Ruusuvuori, K. Liimatainen, S. Albarqouni, B. Mungal, A. George, S. Demirci, N. Navab, S. Watanabe, S. Seno, Y. Takenaka, H. Matsuda, H.A. Phoulady, V. Kovalev, A. Kalinovsky, V. Liauchuk, G. Bueno, M.M. Fernandez-Carrobles, I. Serrano, O. Deniz, D. Racoceanu, R. Venâncio, Diagnostic assessment of deep learning algorithms for detection of lymph node metastases in women with breast cancer, JAMA - J. Am. Med. Assoc. 318 (2017) 2199–2210. https://doi.org/10.1001/jama.2017.14585.

[3]  N. Coudray, P.S. Ocampo, T. Sakellaropoulos, N. Narula, M. Snuderl, D. Fenyö, A.L. Moreira, N. Razavian, A. Tsirigos, Classification and mutation prediction from non–small cell lung cancer histopathology images using deep learning, Nat. Med. (2018). https://doi.org/10.1038/s41591-018-0177-5.

[4]  V. Gulshan, L. Peng, M. Coram, M.C. Stumpe, D. Wu, A. Narayanaswamy, S. Venugopalan, K. Widner, T. Madams, J. Cuadros, R. Kim, R. Raman, P.C. Nelson, J.L. Mega, D.R. Webster, Development and validation of a deep learning algorithm for detection of diabetic retinopathy in retinal fundus photographs, JAMA - J. Am. Med. Assoc. (2016). https://doi.org/10.1001/jama.2016.17216.

[5]  A. Esteva, B. Kuprel, R.A. Novoa, J. Ko, S.M. Swetter, H.M. Blau, S. Thrun, Dermatologist-level classification of skin cancer with deep neural networks, Nature. (2017). https://doi.org/10.1038/nature21056.

[6]  P. Khosravi, E. Kazemi, M. Imielinski, O. Elemento, I. Hajirasouliha, Deep Convolutional

Neural Networks Enable Discrimination of Heterogeneous Digital Pathology Images, EBioMedicine. (2018). https://doi.org/10.1016/j.ebiom.2017.12.026.

[7]     J. Wu, Q. Hu, C. Leng, J. Cheng, Shoot to know what : An application of deep networks on mobile devices, in: 30th AAAI Conf. Artif. Intell. AAAI 2016, 2016.

[8]     J.W. Wei, L.J. Tafe, Y.A. Linnik, L.J. Vaickus, N. Tomita, S. Hassanpour, Pathologist-level classification of histologic patterns on resected lung adenocarcinoma slides with deep neural networks, Sci. Rep. (2019). https://doi.org/10.1038/s41598-019-40041-7.

[9]     Knight Will, The Dark Secret at the Heart of AI - MIT Technology Review, Technologyreview. (2017).

[10]    M. Matheny, S.T. Israni, M. Ahmed, D. Whicher, Artificial Intelligence in Health Care: The Hope, the Hype, the Promise, the Peril, 2019.

[11]    R.R. Selvaraju, M. Cogswell, A. Das, R. Vedantam, D. Parikh, D. Batra, Grad-CAM: Visual Explanations from Deep Networks via Gradient-Based Localization, Int. J. Comput. Vis. (2020). https://doi.org/10.1007/s11263-019-01228-7.

[12]    I. Sim, Mobile devices and health, N. Engl. J. Med. (2019). https://doi.org/10.1056/NEJMra1806949.

[13]    I. Hubara, M. Courbariaux, D. Soudry, R. El-Yaniv, Y. Bengio, Quantized neural networks: Training neural networks with low precision weights and activations, J. Mach. Learn. Res. (2018).

[14]    J. Wu, C. Leng, Y. Wang, Q. Hu, J. Cheng, Quantized convolutional neural networks for mobile devices, in: Proc. IEEE Comput. Soc. Conf. Comput. Vis. Pattern Recognit., 2016. https://doi.org/10.1109/CVPR.2016.521.

[15]    A. Chatterjee, L.R. Varshney, Towards optimal quantization of neural networks, in: IEEE Int. Symp. Inf. Theory - Proc., 2017. https://doi.org/10.1109/ISIT.2017.8006711.

[16]    J. Nalepa, M. Antoniak, M. Myller, P. Ribalta Lorenzo, M. Marcinkiewicz, Towards resource-frugal deep convolutional neural networks for hyperspectral image segmentation, Microprocess. Microsyst. (2020). https://doi.org/10.1016/j.micpro.2020.102994.

[17]    S.J. Frank, A.M. Frank, Salient Slices: Improved Neural Network Training and Performance with Image Entropy, Neural Comput. 32 (2020) 1222–1237. https://doi.org/10.1162/neco_a_01282.

[18]    A.H. Beck, A.R. Sangoi, S. Leung, R.J. Marinelli, T.O. Nielsen, M.J. Van De Vijver, R.B.

West, M. Van De Rijn, D. Koller, Imaging: Systematic analysis of breast cancer morphology uncovers stromal features associated with survival, Sci. Transl. Med. (2011). https://doi.org/10.1126/scitranslmed.3002564.

[19]  D. Justus, J. Brennan, S. Bonner, A.S. McGough, Predicting the Computational Cost of Deep Learning Models, in: Proc. - 2018 IEEE Int. Conf. Big Data, Big Data 2018, 2019. https://doi.org/10.1109/BigData.2018.8622396.

[20]  T. Zhou, S. Ruan, S. Canu, A review: Deep learning for medical image segmentation using multi-modality fusion, Array. (2019). https://doi.org/10.1016/j.array.2019.100004.

| Tile Size | Number of Training Tiles: Entropy / Threshold Gray | Number of Testing Tiles: Entropy / Threshold Gray | Tile Overlap: Entropy / Threshold Gray |
|---|---|---|---|
| 100 | 290,748 / 263,689 | 121,062 / 136,911 | 50% / 43% |
| 150 | 130,161 / 118,766 | 54,670 / 61,455 | 50% / 43% |
| 200 | 73,497 / 66,022 | 30,891 / 34,251 | 50% / 43% |
| 250 | 47,217 / 42,605 | 19,812 / 22,146 | 50% / 43% |
| 300 | 32,802 / 29,380 | 13,839 / 15,322 | 50% / 43% |
| 350 | 24,070 / 21,416 | 10,164 / 11,155 | 50% / 43% |
| 400 | 73,345 / 65,456 | 31,046 / 33,765 | 75% / 71% |
| 450 | 58,423 / 51,872 | 24,774 / 26,730 | 75% / 71% |
| 500 | 46,801 / 39,778 | 19,856 / 20,536 | 75% / 71% |
| 550 | 38,890 / 34,291 | 16,463 / 17,775 | 75% / 71% |
| 600 | 32,359 / 28,906 | 13,713 / 14,940 | 75% / 71% |

| 650 | 27,584 / 24,541 | 11,754 / 12,707 | 75% / 71% |

Supplementary Table 1 – Tile populations for Entropy and Threshold Gray tiles are indicated with corresponding degrees of overlap to achieve these populations. Overlap was increased beginning at 400 × 400 pixel tiles to maintain adequate populations for training and testing.

| Tile Size | XValSet1<br><br>*Entropy train, Entropy test tiles* | XValSet2<br><br>*Entropy train, Entropy test tiles* | XValSet3<br><br>*Entropy train, Entropy test tiles* | XVal Average<br><br>*Entropy train, Entropy test tiles* | XVal Average Variance<br><br>*Entropy train, Entropy test tiles* |
|---|---|---|---|---|---|
| 100 | .92 | .85 | .92 | .90 | 0.06 |
| 150 | .88 | .88 | .92 | .89 | 0.06 |
| 200 | .88 | .85 | .92 | .88 | 0.06 |
| 250 | .92 | .85 | .92 | .90 | 0.06 |
| 300 | .96 | .81 | .88 | .88 | 0.06 |
| 350 | .96 | .85 | .92 | .91 | 0.04 |
| 400 | .92 | .85 | .92 | .90 | 0.06 |
| 450 | .92 | .85 | .92 | .90 | 0.04 |
| 500 | .96 | .85 | .92 | .91 | 0.05 |
| 550 | .92 | .81 | .92 | .88 | 0.04 |
| 600 | .96 | .81 | .92 | .90 | 0.06 |
| 650 | .92 | .81 | .92 | .88 | 0.04 |

Supplementary Table 2 – Prediction accuracies and average variance for three train/test partitions of Entropy tiles.

| Adenocarcinoma | | |
|---|---|---|
| TCGA-05-4244-01Z-00-DX1 | TCGA-05-4426-01Z-00-DX1 | TCGA-56-8305-01Z-00-DX1 |
| TCGA-05-4249-01Z-00-DX1 | TCGA-05-4433-01Z-00-DX1 | TCGA-56-8308-01Z-00-DX1 |
| TCGA-05-4250-01Z-00-DX1 | TCGA-44-3396-01Z-00-DX1 | TCGA-60-2712-01Z-00-DX1 |
| TCGA-05-4397-01Z-00-DX1 | TCGA-44-3918-01Z-00-DX1 | TCGA-63-A5MY-01Z-00-DX1 |
| TCGA-05-4402-01Z-00-DX1 | TCGA-44-4112-01Z-00-DX1 | TCGA-77-7465-01Z-00-DX1 |
| TCGA-05-4427-01Z-00-DX1 | TCGA-44-6148-01Z-00-DX1 | TCGA-94-7033-01Z-00-DX1 |
| TCGA-05-4430-01Z-00-DX1 | TCGA-44-6774-01Z-00-DX1 | TCGA-96-7545-01Z-00-DX1 |
| TCGA-05-4432-01Z-00-DX1 | TCGA-44-7662-01Z-00-DX1 | TCGA-NC-A5HR-01Z-00-DX1 |
| TCGA-05-4434-01Z-00-DX1 | TCGA-44-7669-01Z-00-DX1 | TCGA-56-8623-01Z-00-DX1 |
| TCGA-35-3615-01Z-00-DX1 | TCGA-44-7672-01Z-00-DX1 | TCGA-56-8629-01Z-00-DX1 |
| TCGA-35-4122-01Z-00-DX1 | | TCGA-60-2696-01Z-00-DX1 |
| TCGA-35-4123-01Z-00-DX1 | Squamous | TCGA-68-8250-01Z-00-DX1 |
| TCGA-38-4625-01Z-00-DX1 | TCGA-18-5595-01Z-00-DX1 | TCGA-77-7140-01Z-00-DX1 |
| TCGA-38-4626-01Z-00-DX1 | TCGA-21-1070-01Z-00-DX1 | TCGA-77-8138-01Z-00-DX1 |
| TCGA-38-4631-01Z-00-DX1 | TCGA-21-1077-01Z-00-DX1 | TCGA-77-A5GA-01Z-00-DX1 |
| TCGA-38-6178-01Z-00-DX1 | TCGA-21-1078-01Z-00-DX1 | TCGA-85-A4PA-01Z-00-DX1 |
| TCGA-44-2656-01Z-00-DX1 | TCGA-21-1081-01Z-00-DX1 | TCGA-85-A510-01Z-00-DX1 |
| TCGA-44-2659-01Z-00-DX1 | TCGA-21-1083-01Z-00-DX1 | TCGA-94-7557-01Z-00-DX1 |
| TCGA-44-2661-01Z-00-DX1 | TCGA-21-A5DI-01Z-00-DX1 | TCGA-96-A4JK-01Z-00-DX1 |
| TCGA-44-2662-01Z-00-DX1 | TCGA-22-1017-01Z-00-DX1 | TCGA-98-8023-01Z-00-DX1 |
| TCGA-44-3398-01Z-00-DX1 | TCGA-22-5471-01Z-00-DX1 | TCGA-J1-A4AH-01Z-00-DX1 |
| TCGA-44-5644-01Z-00-DX1 | TCGA-22-5472-01Z-00-DX1 | |
| TCGA-44-5645-01Z-00-DX1 | TCGA-22-5474-01Z-00-DX1 | |
| TCGA-44-6145-01Z-00-DX1 | TCGA-34-2605-01Z-00-DX1 | |
| TCGA-44-6146-01Z-00-DX1 | TCGA-37-4141-01Z-00-DX1 | |
| TCGA-44-6776-01Z-00-DX1 | TCGA-39-5039-01Z-00-DX1 | |
| TCGA-44-6777-01Z-00-DX1 | TCGA-43-7657-01Z-00-DX1 | |
| TCGA-44-7659-01Z-00-DX1 | TCGA-43-7658-01Z-00-DX1 | |
| TCGA-44-8119-01Z-00-DX1 | TCGA-43-A474-01Z-00-DX1 | |
| TCGA-05-4396-01Z-00-DX1 | TCGA-51-6867-01Z-00-DX1 | |
| TCGA-05-4420-01Z-00-DX1 | TCGA-56-7221-01Z-00-DX1 | |
| TCGA-05-4424-01Z-00-DX1 | TCGA-56-7222-01Z-00-DX1 | |
| | TCGA-56-8201-01Z-00-DX1 | |

| Out-of-Sample Adenocarcinoma | TCGA-73-4670-01Z-00-DX1 | TCGA-39-5011-01Z-00-DX1 |
|---|---|---|
| TCGA-44-2659-01Z-00-DX1 | TCGA-99-8025-01Z-00-DX1 | TCGA-39-5040-01Z-00-DX1 |
| TCGA-55-1595-01Z-00-DX1 | | TCGA-60-2714-01Z-00-DX1 |
| TCGA-55-7725-01Z-00-DX1 | Out-of-Sample Squamous | TCGA-92-7340-01Z-00-DX1 |
| TCGA-66-2785-01Z-00-DX1 | TCGA-21-5783-01Z-00-DX1 | |
| | TCGA-37-4130-01Z-00-DX1 | |

Supplementary Table 3 – Primary and out-of-sample datasets

## Conflict of Interest Statement

There are no financial or personal relationships with other people or organisations that could inappropriately influence or bias this work.